\documentclass[10pt, a4paper]{article}
\usepackage{lrec}
\usepackage{multibib}
\newcites{languageresource}{Language Resources}
\usepackage{graphicx}
\usepackage{tabularx}
\usepackage{soul}
\usepackage{listings}
\usepackage{mathtools}
\usepackage{amssymb}

\usepackage{epstopdf}
\usepackage[utf8]{inputenc}

\usepackage{hyperref}
\usepackage{xstring}
\usepackage{xifthen}

\usepackage{color}

\title{Multilingual enrichment of disease biomedical ontologies} 

\name{Léo Bouscarrat$^{1,2}$, Antoine Bonnefoy$^{1}$, Cécile Capponi$^{2}$, Carlos Ramisch$^{2}$}

\address{$^1$EURA NOVA, Marseille, France \\
         $^2$Aix Marseille Univ, Université de Toulon, CNRS, LIS, Marseille, France \\
         \{leo.bouscarrat, antoine.bonnefoy\}@euranova.eu\\
         \{leo.bouscarrat, cecile.capponi, carlos.ramisch\}@lis-lab.fr\\}

\abstract{
Translating biomedical ontologies is an important challenge, but doing it manually requires much time and money. We study the possibility to use open-source knowledge bases to translate biomedical ontologies. We focus on two aspects: coverage and quality. We look at the coverage of two biomedical ontologies focusing on diseases with respect to Wikidata for 9 European languages (Czech, Dutch, English, French, German, Italian, Polish, Portuguese and Spanish) for both ontologies, plus Arabic, Chinese and Russian for the second one. We first use direct links between Wikidata and the studied ontologies and then use second-order links by going through other intermediate ontologies.
We then compare the quality of the translations obtained thanks to Wikidata with a commercial machine translation tool, here Google Cloud Translation.
\\ \newline  }

\begin{document}

\maketitleabstract

\begin{figure*}[ht!]
\centering
\includegraphics[width=0.38\textwidth]{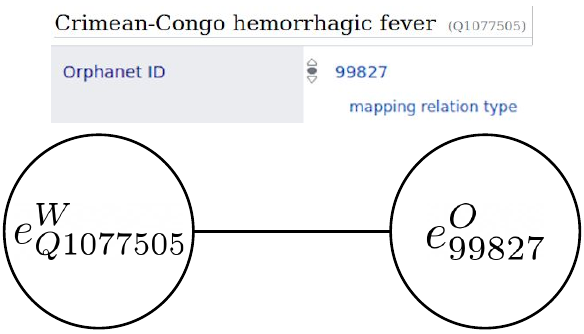}
\includegraphics[width=0.61\textwidth]{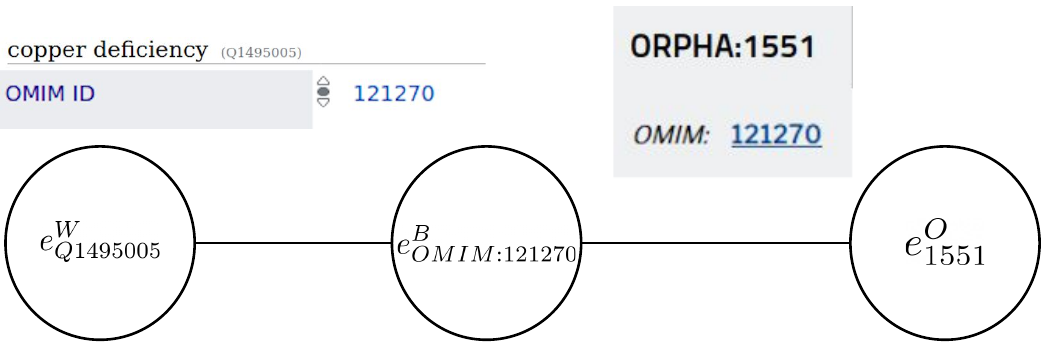}
\caption{Example of first-order link (left) and second-order link (right)}
\label{fig:links}
\end{figure*}

\section{Introduction}

Biomedical ontologies, like Orphanet \cite{Orphanet}, play an important role in many downstream tasks \cite{andronis_literaturemining_2011,li_repurposigng_2015,phan2017ontology}, especially in natural language processing \cite{maldonado2017deep,nayel_integrating_2019}. Today either the vast majority of these ontologies are only available in English or their restrictive licenses reduce the scope of their usage. There is nowadays a real focus on reducing the prominence of English, thus on working on less-resourced languages. To do so, there is a need for resources in other languages, but the creation of such resources is time and money consuming.

At the same time, the Internet is also a source of incredible projects aiming to gather a maximum of knowledge in a maximum of languages. One of them is the collaborative encyclopedia Wikipedia, opened in 2001, which currently exists in more than 300 languages. As it contains mainly plain text, it is hard to use it as a resource as is. However, several knowledge bases have been built from it: DBpedia \cite{lehmann2015dbpedia} and Wikidata \cite{vrandevcic2014wikidata}. The main difference between these two knowledge graphs is the update process: while Wikidata is manually updated by users, DBpedia extracts its information directly from Wikipedia. Compared to biomedical ontologies they are structured using less expressive formalisms and they gather information about a larger domain. They are open-source, thus can be used for any downstream tasks. For each entity they have a preferred label, but sometimes also alternative labels that can be used as synonyms. For example, the entity \textit{Q574227} in Wikidata has the preferred label \textit{2q37 monosomy} in English along with the alternative labels in English: \textit{Albright Hereditary Osteodystrophy-Like Syndrome} and \textit{Brachydactyly Mental Retardation Syndrome}. Moreover, entities in these two knowledge bases also have translations in several languages. For example, the entity \textit{Q574227} in Wikidata has the preferred label \textit{2q37 monosomy} in English and the preferred label \textit{Zespół delecji 2q37} in Polish. They also feature some links between their own entities and entities in external biomedical ontologies. For example, the entity \textit{Q574227} in Wikidata has a property \textit{Orphanet ID} (\textit{P1550}) with the value \textit{1001}.

By using both kinds of resources, biomedical ontologies and open-source knowledge bases, we could partially enrich biomedical ontologies in languages other than English. As links between the entities of these resources are already existing, we expect good quality. To further enrich them we could even look at second-order links since many biomedical ontologies also contain some links to other ontologies.


The goal of this work is twofold:
\begin{itemize}
    \item to study the coverage of such open-source collaborative knowledge graphs compared to biomedical ontologies,
    \item to study the quality of the translations using first- and second-order links and comparing this quality with the quality obtained by machine translation tools.
\end{itemize}

This paper is part of a long-term project whose goal is to work on multilingual disease extraction from news with strategies based on dictionary expansion. Consequently, we need a multilingual vocabulary with diseases which are normalized with respect to an ontology. Thus, we focus on one kind of biomedical ontologies, that is, ontologies about diseases.

\section{Resources and Related Work}

There has already been some work trying to use open-source knowledge bases to translate biomedical ontologies.

\newcite{bretschneider_corpus-based_2014} obtain a German-English medical dictionary using DBPedia. The goal is to perform information extraction from a German biomedical corpus. They could not directly use the RadLex ontology \cite{langlotz2006radlex} as it is only available in English. So, they first extract term candidates in their German corpus. Then, they try to match the candidates with the pairs in their German-English dictionary. If a candidate is in the dictionary, they use the translation to match with the RadLex ontology. Finally, this term candidate alongside with the match in the RadLex ontology is processed by a human to validate the matching.

\newcite{alba_multi-lingual_2017} create a language-independent method to maintain up-to-date ontologies by extracting new instances from text. This method is based on a human-in-the-loop who helps tuning scores and thresholds for the extraction. Their method requires some ``contexts'' to start finding new entities to add to the ontology. To bootstrap the contexts, they can either ask a human to annotate some data or use an oracle made by the dictionary extracted from the DBpedia and Wikidata using word matching on the corpus. They then look for good candidates, i.e., a set of words surrounding an item, by looking for elements in similar contexts to the one found using the bootstrapping. Then, a human-in-the-loop validates the newly found entities, adding them to the dictionary if they are correct, or down-voting the context if they are not relevant entities.

\newcite{hailu_ontology_2014} work on the translation of the Gene Ontology from English to German and compare three different approaches: DBpedia, the Google Translate API without context, and the Google Translate API with context. To find the terms in DBpedia they use keyword-based search. After a human evaluation, they find that translations obtained with DBpedia have the lowest coverage (only 25\%) and quality compared to those obtained with Google Translate API. However, to compare the quality of the different methods they only use the translation of 75 terms obtained with DBpedia compared to 1,000 with Google Translate API. They also note that synonyms could be a useful tool for machine translation and that using keyword-based exact match query to match the two sources could explain the low coverage.

\newcite{silva_ontology-based_2015} compare three methods to translate SNOMED CT from English to Portuguese: DBpedia, ICD-9 and Google Translate. To verify the quality of the different approaches they use the CPARA ontology which has been hand-mapped to SNOMED CT. It is composed of 191 terms and focused on allergies and adverse reactions. They detect coverage of 10\% with the ICD-9, 37\% with DBpedia and 100\% with Google Translate. To compare the quality of their translations they use the Jaro Similarity \cite{jaro1989advances}.

We elaborate on these ideas by adding some elements. First of all, compared to \newcite{hailu_ontology_2014} and \newcite{silva_ontology-based_2015}, we use already existing properties to perform the matching between the biomedical ontology and the knowledge graph, which should improve the quality with regard to the previous works. We also go further than these first-order links and explore the possibility of using second-order links to improve the coverage of the mappings between the sources. Compared to the same works, we also present a more complete study, \newcite{hailu_ontology_2014} only evaluate on 75 terms and \newcite{silva_ontology-based_2015} on 191 terms. We compare the coverage and quality of the entire biomedical ontology containing 10,444 terms. Furthermore, as we want to use the result of this work for biomedical entity recognition, synonyms of entities are really important for recall and also for normalisation, thus we also quantify the difference of quantity of synonyms between the original biomedical ontology and those found with Wikidata.

In this work, as we focus on diseases, we use a free dataset extracted from Orphanet \cite{Orphanet} to perform the evaluation. Orphanet is a resource built to gather and improve knowledge about rare diseases. Through Orphadata \cite{Orphadata}, free datasets of aggregated data are updated monthly. One of them is about rare diseases, including cross-references to other ontologies. The Orphadata dataset contains the translation of 10,444 entities for English, French, German, Spanish, Dutch, Italian, Portuguese, 10,418 entities in Polish and 9,323 in Czech. All the translations have been validated by experts, thus can be used as a gold standard for multilingual ontology enrichment. One issue of this dataset is that rare diseases are, by definition, not well known. Therefore, one may expect a lower coverage than a less focused dataset; thus we propose to also measure the coverage of another dataset, Disease Ontology \cite{schriml2019human}. However we cannot use it to evaluate the translation task as it does not contain translations.

As an external knowledge base, we use Wikidata. It has many links to external ontologies, especially links to biomedical ontologies such as \textit{wdt:P1550} for Orphanet, \textit{wdt:P699} for Disease Ontology, and \textit{wdt:P492} for the Online Mendelian Inheritance in Man (OMIM). It is also important to note that, over the 9 languages we studied, only the Czech Wikipedia has less than 1,000,000 articles. This information can be used as a proxy for the completeness of the information in each language on Wikidata. We prefer it over DBpedia as we find it easier to use, especially to find the properties.

As a machine translation tool, we use Google Cloud Translation. It is a paying service offered by Google Cloud.

\section{Methods and Experiments}

In this section, we first define the notations used in this paper, then we describe how we extract the first- and second-order links from our sources. Afterwards, we describe how we perform machine translation. The evaluation metrics are subsequently explained and finally we describe our evaluation protocol.

\subsection{Definition and Notations}

\newcommand{\getLabel}[2]{\operatorname{L}_{#1}(#2)}
\newcommand{\getAlt}[2]{\operatorname{\mathcal{L}}_{#1}(#2)}
\newcommand{\neighbours}[1]{\operatorname{\mathcal{V}}(#1)}
\newcommand{\neighboursWiki}[1]{\operatorname{\mathcal{W}}(#1)}
\newcommand{\jaroSim}[2]{\operatorname{J}(#1,#2)}
\newcommand{\jaroMax}[2]{\operatorname{\mathcal{J}}_{\max}\left(#1, #2\right)}
\newcommand{\measureOne}[3][]{\operatorname{\mathcal{M}_pl}\ifthenelse{\equal{#1}{}}{}{\left(#1, #2, #3\right)}}
\newcommand{\measureTwo}[3]{\operatorname{\mathcal{M}_bl}\ifthenelse{\equal{#1}{}}{}{\left(#1, #2, #3\right)}}
\newcommand{\measureThree}[3]{\operatorname{\mathcal{M}_mbl}\ifthenelse{\equal{#1}{}}{}{\left(#1, #2, #3\right)}}
\newcommand{\measureFour}[3]{\operatorname{\mathcal{M}_Mbl}\ifthenelse{\equal{#1}{}}{}{\left(#1, #2, #3\right)}}

We define:
\begin{itemize}
    \item $e_i^S$ as an entity in the source knowledge base $S$, $S \in [O, W, B]$ where $O$ is Orphanet, $W$ is WikiData and $B$ are all the other external biomedical ontologies used. An entity is either a concept in an ontology or in a knowledge graph. \\
    \item $E^S = \{e_i^S\}_{i = 1 ... |E^S|}$ is the set of all the entities in the source $S$. \\
    \item $E = E^O \cup E^W \cup E^B$ is the set of all the entities in all the sources.\\
    \item $\getLabel{l}{e}$ is the preferred label of the entity $e$ in the language $l$, or $\emptyset$ if there is no label in this language. \\
    \item $\getAlt{l}{e}$ represents all the possible labels of the entity $e$ in the language $l$ or $\emptyset$ if there is no label in this language. Furthermore, $\getLabel{l}{e} \in \getAlt{l}{e}$ \\
    \item $T$ is a set of links, such that $t \in T$ with $ t = (e_i^s, e_j^{s'}), s \neq s'$. \\
    \item $G = (E, T)$ is an undirected graph. \\
    \item $\neighbours{e_i} = \{e_j \in E | \exists t \in T, t = (e_i, e_j)\}$, defines the set of all the neighbours of the entity $e_i$. \\
    \item $\neighboursWiki{e} = \{v \in \neighbours{e} | v \in W\}$, defines the set of all the neighbours that are in Wikidata of the entity $e$. \\
    \item $MT(\{s_1, ..., s_n\}, l)$ is a function that returns the labels $\{s_1, ...,s_n\}$ translated from English to the language $l$ thanks to Google Cloud Translation.
\end{itemize}


\subsection{Gathering Links between Entities}
\label{subsection:gathering_links}

\subsubsection{First-Order Links}

The first step of our method consists in gathering all the information about the sources. To obtain the gold translations, we use Orphadata. We collected all the JSON files from their website\footnote{\url{http://www.orphadata.org/cgi-bin/rare\_free.html}} on  January 15, 2020. We extract the OrphaNumber, the Name, the SynonymList and the ExternalReferenceList of each element in the files. 

For WikiData we use the SPARQL endpoint\footnote{\url{https://query.wikidata.org/sparql} can be queried with the interface \url{https://query.wikidata.org/}}. We query all the entities having a property OrphaNumber \textit{wdt:P1550}, and, for these entities, we obtain all their preferred labels (\textit{rdfs:label}) and synonyms (\textit{skos:altLabel}), corresponding to $E_i^O$ in the 9 European languages included in Orphanet. The base aggregator of the synonyms uses a comma to separate them. In our case, this error-prone because the comma can also be part of the label, for example one of the alternative label of the entity \textit{Q55786560} is \textit{49, XXXYY syndrome}. We needed to concatenate the synonyms with another symbol\footnote{We made a package to extract entities from Wikidata: \url{https://github.com/euranova/wikidata_property_extraction}}. Thanks to the property which gives the Orphanumber of the related entity in Orphanet we can create links $t = (e^O, e^W)$ between an entity $e_i^W$ in Wikidata and and entity $e_i^O$ in Orphanet. The queries have been made on April 01, 2020.

The mapping is then trivial, as we have the OrphaNumber in the two sources. On the left of Figure~\ref{fig:links} we can see that the entity \textit{Q1077505} in Wikidata has a property \textit{Orphanet ID} with the value \textit{99827}, thus we can create $t = (Q1077505^W, 99827^O)$. Nonetheless, the mapping is not always unary, because several Wikidata entities can be linked to the same Orphanet entity.

Formally, the set of Orphanet entities with at least one first-order link is: $$E^F = \{e \in E^O| \exists w \in W, (e, w) \in T \}$$

\subsubsection{Second-Order Links}

Orphanet provides some external references to auxiliary ontologies. We add these references to our graph: $t = (e^O, e^B) \in T$. Even if there are already first-order links between Orphanet and Wikidata, we cannot ensure that all the entities are linked. To improve the coverage of translations, we can use second-order links, creating an indirect link when entities from Wikidata and Orphanet are linked to the same entity in a third external source $B$. For example, on the right of Figure \ref{fig:links}, we extract the link between the entity \textit{Q1495005} of Wikidata and the entity \textit{121270} of OMIM. We also extract from Orphanet that the entity \textit{1551} of Orphanet is link to the same entity of OMIM. Therefore, as a second-order relation, the entity \textit{Q1495005} of Wikidata and the entity \textit{1551} of Orphanet are linked.

The objective is to find some links $t' = (e^W, e^B)$ where $\exists v \in \neighbours{e^B}$ and $v \in E^O$. Consequently, we are looking for links between entities from Wikidata and the external biomedical ontologies, whenever the entity in the external biomedical ontology already has a link with an entity in Orphanet.

For that purpose, we extract all the links between Wikidata and the external biomedical ontologies in the same fashion as from Orphanet, using the appropriate Wikidata properties. In the previous example, we create  links $(Q1495005^W, OMIM:121270^B) \in T$ and $(1551^O, OMIM:121270^B) \in T$. 

We can now map Wikidata and Orphanet using second-order links. This set of links is denoted as:

\begin{equation*}
\begin{multlined}[t]
C = \{e \in E^O | \exists (w, b) \in E^W \times E^B, \\
(e, b) \in T, (w, b) \in T\}
\end{multlined}
\end{equation*}

We also define the set of all the second-order linked Wikipedia entities of a specific Orphanet entity:

\begin{equation*}
\begin{multlined}[t]
\mathcal{C}(e^O) = \{w \in E^W | \exists b \in E^B, (e, b) \in T, (w, b) \in T\}
\end{multlined}
\end{equation*}

\subsection{Machine Translation}

We use Google Cloud Translation as a machine translation tool to translate the labels of the ontology from English to a target language. As we want to have the same entities in the test set as for Wikidata, for each language we only translate the Orphanet entities which have at least one first-order link to an entity in Wikidata with a label in the target language. So for an entity $e$, for the language $l$ the output of Google Cloud Translation is: $$MT(\getAlt{en}{e}, l)$$

\subsection{Definition of Evaluation Metrics}

In this section, we define the different evaluation metrics that are used to evaluate the efficiency of the method.

\subsubsection{Coverage Metric}

To estimate the coverage of Wikipedia on a biomedical ontology we use the following metric:

\begin{equation*}
    Coverage(E_1, E_2, l) = \frac{|\{e \in E_1| \getLabel{l}{e} \neq \emptyset \}|}{|\{e' \in E_2| \getLabel{l}{e'} \neq \emptyset \}|}
\end{equation*}
where $E_1$ and $E_2$ are sets of entities.

\subsubsection{Jaro Similarity and n-ary Jaro}

In order to evaluate the quality of the translations, we follow  \newcite{silva_ontology-based_2015} choosing the Jaro similarity, which is a type of edit distance. We made this choice as we are looking at entities. Whereas other measures such as BLEU \cite{papineni2002bleu} are widely used for translation tasks, they have been designed for full sentences instead of relatively short ontology labels. 
The Jaro Similarity is defined as:

\begin{equation*}
\jaroSim{s}{s'} = \frac{1}{3}\left(\frac{m}{\mid s \mid} + \frac{m}{\mid s' \mid} + \frac{m - t}{m}\right) s, s' \in \{a, ..., z\}^* 
\end{equation*}

with $s$ and $s'$ two strings, 
$\mid s \mid$ the length of $s$,
$t$ is half the number of transpositions,
$m$ the number of \textit{matching characters}.  Two characters from $s$ and $s'$ are \textit{matching} if they are the same and not further than $\frac{max(\mid s \mid, \mid s' \mid)}{2} - 1$.
The Jaro Similarity ranges between 0 and 1, where the score is 1 when the two strings are the same.

However, since one Orphanet entity may have several neighbour Wikidata entities, we cannot use the Jaro similarity directly. We choose to use the $\max$, for considering the quality of the closest entity:

\begin{equation*}
    \jaroMax{s}{[s_1, ..., s_n]} = \underset{s' \in [s_1, ..., s_n]}{\max}\jaroSim{s}{s'}
\end{equation*}

\subsubsection{Quality Metrics}
\label{subsubsection:quality_metrics}

From assessing the quality of the translations, we create 4 different measures with different goals. For each entity in each language, there is a preferred label $\getLabel{l}{e}$ and a list of all the possible labels $\getAlt{l}{e}$. All of the metrics range between 0 and 1, the higher the better.

\begin{equation*}
    \measureOne[e]{[e_1, ..., e_n]}{l} = \jaroMax{\getLabel{l}{e}}{[\getLabel{l}{e_1}, .., \getLabel{l}{e_n}]}
\end{equation*}

\begin{equation*}
    \measureTwo{e}{[e_1, ..., e_n]}{l} = \jaroMax{\getLabel{l}{e}}{\bigcup\limits_{i=1}^{n}\getAlt{l}{e_i}}
\end{equation*}

\begin{equation*}
    \measureThree{e}{[e_1, ..., e_n]}{l} = \underset{s \in \getAlt{l}{e}}{mean}\jaroMax{s}{\bigcup\limits_{i=1}^{n}\getAlt{l}{e_i}}
\end{equation*}

\begin{equation*}
    \measureFour{e}{[e_1, ..., e_n]}{l} =\underset{s \in \getAlt{l}{e}}{\max}\jaroMax{s}{\bigcup\limits_{i=1}^{n}\getAlt{l}{e_i}}
\end{equation*}

$\measureOne{}{}{}$, for principal label, compares the preferred labels from Orphanet and Wikidata. This number is expected to be high, but as there is no reason that Wikidata and Orphanet use the same preferred label, we do not expect it to be the highest score. Nonetheless, as Wikidata is a collaborative platform, a score of 1 on a high number of entities in a different language could also indicate that the translations come from Orphanet.

$\measureTwo{}{}{}$, for best label, compares the preferred label from Orphanet against all the labels in Wikidata. The goal here is to verify that the preferred label of Orphanet is available in Wikidata.

$\measureThree{}{}{}$, for mean best label, takes the average of the similarity of one label in Orphanet against all the labels in Wikidata. This score can be seen as a completeness score, it evaluates the ability of finding all the labels of Orphanet in Wikidata.

$\measureFour{}{}{}$, for max best label, takes the maximum of the similarity of one label in Orphanet against all the labels in Wikidata. The question behind this metric is: Do we have at least one label in common between Orphanet and Wikidata? A low score here could mean that the relation is erroneous. We expect a score close to 1 here.

We used the same measures for the machine-translated dataset, however, the difference between $\measureOne{}{}{}$ and $\measureTwo{}{}{}$ is expected to be smaller, as we are sure that the preferred label from the translated dataset is the translation of the preferred label from Orphanet.

To obtain a score for these measures on the entire dataset, we compute the average of the scores over all Orphanet entities.

\begin{table*}[ht]
\begin{center}
\begin{tabular}{|c|c|c|c|c|c|c|c|c|c|c|c|c|c|c|c|c|}

      \hline
       & \multicolumn{3}{c|}{$\measureOne{}{}{}$} &\multicolumn{3}{c|}{$\measureTwo{}{}{}$} & \multicolumn{3}{c|}{$\measureThree{}{}{}$} & \multicolumn{3}{c|}{$\measureFour{}{}{}$} \\
      \hline
      Lang & 1st W & 1+2nd W & GCT & 1st W & 1+2nd W & GCT & 1st W & 1+2nd W & GCT & 1st W & 1+2nd W & GCT\\
      \hline
      EN & 85.5 & \textbf{87.5} & N/A & 91.5 & \textbf{92.1} & N/A & \textbf{84.1} & 80.5 & N/A & \textbf{97.3} & 96.6 & N/A \\
      \hline
      FR & 85.3 & 82.4 & \textbf{89.8} & 87.4 & 84.2 & \textbf{90.5} & 75.7 & 69.3 & \textbf{90.1} & 94.1 & 89.1 & \textbf{97.7} \\
      \hline
      DE & 77.1 & 67.8 & \textbf{80.5} & 79.1 & 70.3 & \textbf{81.6} & 67.5 & 60.9 & \textbf{83.4} & 88.7 & 79.0 & \textbf{95.4} \\
      \hline
      ES & 81.3 & 70.1 & \textbf{92.5} & 84.4 & 73.0 & \textbf{93.0} & 68.7 & 58.4 & \textbf{90.2} & 91.7 & 89.1 & \textbf{98.3} \\
      \hline
      PL & 78.0 & 63.8 & \textbf{82.0} & 82.0 & 61.3 & \textbf{83.2} & 66.6 & 55.9 & \textbf{85.0} & 90.7 & 77.3 & \textbf{95.7} \\
      \hline
      IT & 79.4 & 66.7 & \textbf{88.4} & 82.4 & 68.8 & \textbf{89.5} & 69.1 & 58.5 & \textbf{88.1} & 90.5 & 77.4 & \textbf{97.2} \\
      \hline
      PT & 79.9 & 64.9 & \textbf{83.6} & 82.1 & 66.5 & \textbf{87.6} & \textbf{73.7} & 60.8 & 68.4 & 93.5 & 83.5 & \textbf{93.3} \\
      \hline
      NL & 72.9 & 59.1  & \textbf{88.0} & 75.6 & 60.9 & \textbf{88.7} & 65.8 & 55.1 & \textbf{89.9} & 86.5 & 71.4 & \textbf{97.2} \\
      \hline
      CS & 76.3 & 52.8 & \textbf{81.9} & 79.1 & 54.9 & \textbf{83.3} & 67.5 & 52.3 & \textbf{85.4} & 88.7 & 68.8 & \textbf{95.3} \\
      \hline

\end{tabular}
\caption{Scores of the different methods with the different metrics in function of the languages. 1st W represents the quality of the first-order links with Wikidata, 1+2nd W the first and second-order links, and GCT the translations obtained by Google Cloud Translation.}
\label{table:quality}
\end{center}
\end{table*}

\subsection{Protocol}
\label{subsection:Protocol}

The first step of our experiments is the extraction of first-order and second-order links from Wikidata and Orphanet as explained in \ref{subsection:gathering_links}. Once these links are available, we study them, starting with their coverage. To evaluate the coverage of Wikidata for each language, we compute $Coverage(E^F, E^O, l)$ for the 9 languages. We also compute $Coverage(C, E^O, l)$ for second-order links. As Orphanet is focused on rare diseases, we do not expect a high coverage in Wikidata. To verify this hypothesis, we do the same evaluation on the Disease Ontology, which does not focus on rare diseases.

Then, we study the quality of the different methods. We apply the 4 quality metrics defined in \ref{subsubsection:quality_metrics} for each language on each method:
\begin{itemize}
    \item First-order links: $\underset{e^O \in E^F}{mean}(\mathcal{M}(e^O, \neighboursWiki{e^O}, l)$
    \item Second-order links: $\underset{e^O \in C}{mean}(\mathcal{M}(e^O, \mathcal{C}(e^O), l)$
    \item Machine translation: $\underset{e^O \in E^F}{mean}(\mathcal{M}(e^O, MT(\getAlt{e^O}{l}, l), l)$
\end{itemize}
Finally, we look at the number of labels we can obtain for both sources. 
\begin{itemize}
    \item Orphanet: $\underset{e \in E^F}{mean}|\getAlt{l}{e}|$
    \item Wikidata: $\underset{e \in E^F}{mean}\underset{w \in \neighboursWiki{e}}{\sum}|\getAlt{l}{w}|$
    \item GCT: $\underset{e \in E^F}{mean}|MT(\getAlt{en}{e}, l)|$
\end{itemize}
The number of synonyms of an entity $e$ in a language $l$ is: $|\getAlt{l}{e}|$, and we also remove the duplicates. We then average this over all the entities which are in a first-order link and in Wikidata and Orphanet.

\section{Results}

In this part, we first present the results on the coverage of Wikipedia on Orphanet, then we present the quality of the translation. Afterwards, we show results about the number of synonyms in both sources and finally we discuss these results.\footnote{The results can be reproduced with this code: \url{https://github.com/euranova/orphanet_translation}}

\subsection{Coverage}

\subsubsection{Orphanet}

First, we evaluate the coverage for each language, i.e., the percentage of entities in Orphanet which have at least one translation in Wikidata.

The Orphadata dataset contains translations of English, French, German, Spanish, Dutch, Italian, Portuguese, Polish and Czech. For Wikidata, the results depend on the language as not all the entities have translations in every language.

\begin{table}[!h]
\begin{center}
\begin{tabular}{|l|c|c|}

      \hline
      Language & Orphanet & Wikidata (\%)\\
      \hline
      English & 10,444 & 8,870 (84.9\%) \\
      \hline
      French & 10,444 & 5,038 (48.2\%) \\
      \hline
      German & 10,444 & 1,946 (18.6\%) \\
      \hline
      Spanish & 10,444 & 1,565 (15.0\%) \\
      \hline
      Polish & 10,171 & 1,329 (13.1\%) \\
      \hline
      Italian & 10,444 & 1,175 (11.3\%) \\
      \hline
      Portuguese & 10,444 & 921 (8.8\%) \\
      \hline
      Dutch & 10,444 & 888 (8.5\%) \\
      \hline
      Czech & 9,323 & 452 (4.8\%) \\
      \hline

\end{tabular}
\caption{Number of translated entities in Orphanet and number of Orphanet entities having at least one translation in Wikidata with first-order links. The percentage of coverage is shown in parentheses.}
\label{table:coveragefirstorder}
 \end{center}
\end{table}

As we can see in Table~\ref{table:coveragefirstorder} that coverage depends on the language. The coverage of English gives us the amount of entities from Orphanet having at least one link with Wikidata. Here, we have 84.9\% of the entities which are already linked to at least one entity in Wikidata. It means that the property of the OrphaNumber is widely used. We can also note that the French Wikidata seems to carry more information about rare diseases than the German Wikipedia. Indeed French and German Wikipedias have approximately the same global size\footnote{As of the 6th February 2020: \url{https://meta.wikimedia.org/wiki/List_of_Wikipedias}}, but the German Wikidata contains much less information about rare diseases.

\begin{table}[!h]
\begin{center}
\begin{tabular}{|l|c|c|}
      \hline
      Language & Cov 1st (\%) & Cov 1st+2nd (\%) \\
      \hline
      English & 8,870 (84.9\%) & 9,317 (89.2\%) \\
      \hline
      French & 5,038 (48.2\%) & 7,922 (75.9\%) \\
      \hline
      German & 1,946 (18.6\%) & 6,350 (60.8\%) \\
      \hline
      Spanish & 1,565 (15.0\%) & 6,122 (58.6\%) \\
      \hline
      Polish & 1,329 (13.1\%) & 5,797 (57.0\%) \\
      \hline
      Italian & 1,175 (11.3\%) & 5,715 (54.7\%)\\
      \hline
      Portuguese & 921 (8.8\%) & 5,016 (48.0\%) \\
      \hline
      Dutch & 888 (8.5\%) & 5,081 (48.6\%) \\
      \hline
      Czech & 452 (4.8\%) & 3,180 (34.1\%) \\
      \hline
\end{tabular}
\caption{Coverage in terms of number and percentage of entities in Wikidata linked to Orphanet using first-order links (Cov 1st) and first- plus second-order links (Cov 1st+2nd).}
\label{table:coveragesecondorder}
\end{center}
\end{table}

The next question is the quantity of new links we can obtain by gathering second-order links.

Table \ref{table:coveragesecondorder} shows that the second-order links improve the coverage. For English, the improvement is small. Thus, for all the other languages, second-order links really help to increase the coverage. It seems to be a good help for average-resourced languages. We have used ICD-10, Medical Subject Heading (MeSH), Online Mendelian Inheritance in Man (OMIM), and, Unified Medical Language System (UMLS) as auxiliary ontologies.


\subsubsection{Disease Ontology}

Even if the coverage for Orphanet in English is already high, Orphanet is focused on rare diseases, which is really specific. This specificity could have an impact on the coverage as Wikidata is not made by experts. To verify if the specificity of this ontology has an influence on coverage, we have also looked at another biomedical ontology on diseases, Disease Ontology. It is also about diseases but does not focus on rare disease. Thus, this difference in generality is expected to have an impact on the coverage.

The Disease Ontology contains 12,171 concepts. We plan to use it for future works on other languages: Arabic, Russian and Chinese. These three languages also have Wikipedias with more than 1,000,000 articles on which we could rely.

\begin{table}[!h]
\center
\begin{tabular}{|l|c|}

      \hline
      Language & Wikidata (\%)\\
      \hline
      English & 11,833 (97.2\%) \\
      \hline
      French & 7,156 (58.8\%) \\
      \hline
      Spanish & 3,178 (26.1\%) \\
      \hline
      Arabic & 2,507 (20.6\%) \\
      \hline
      German & 2,500 (20.5\%) \\
      \hline
      Italian & 2,098 (17.2\%) \\
      \hline
      Polish & 1,869 (15.3\%) \\
      \hline
      Chinese & 1,789 (14.7\%) \\
      \hline
      Portuguese & 1,748 (14.3\%) \\
      \hline
      Russian & 1,706 (14.0\%) \\
      \hline
      Dutch & 1,650 (13.6\%) \\
      \hline
      Czech & 1,001 (8.2\%) \\
      \hline

\end{tabular}
\caption{Number of entities in Disease Ontology translated, number of Disease Ontology entities having at least one translation in Wikidata with first order links and the percentage of coverage.}
\label{table:coveragedo}
\center
\end{table}

As expected, this less expert ontology seems to have better coverage than Orphanet. Table \ref{table:coveragedo} shows that, even if the coverage for all the languages is better than for Orphanet, the difference is not the same for all the languages. Especially, Spanish has a coverage in Disease Ontology superior to that in Orphadata by more than 11\%. We do not have an explanation for these differences.
 
We do not compute the second-order links for Disease Ontology because 97.2\% of the Orphanet entities are already linked using first-order links.

\subsection{Quality}

The next question concerns the quality of the translations obtained. We can expect high-quality translations from Google Cloud Translation, but to what extent? We also want to compare the quality of translations obtained from Wikidata using first-order and second-order links. 
The ontology we use is heavily linked directly to Wikidata, but this is not the case for all the ontologies. For ontologies with lower first-order coverage, one could expect higher increase of the second-order coverage as observed in Table \ref{table:coveragesecondorder}.

The first line of Table \ref{table:quality} shows the matching between the English labels of the entities of Orphanet and Wikidata. $\measureTwo{}{}{}$ and $\measureFour{}{}{}$ are interesting here as they can be used as an indicator of a good match. A score of 1 means that one of the labels of Wikidata is the same as the preferred label from Orphanet ($\measureTwo{}{}{}$) or one of the labels from Orphanet ($\measureFour{}{}{}$). Considering that the scores are close to 1, the matching seems to be good.

In Table \ref{table:quality} we can see that Google Cloud Translation gives the best translations when evaluated with the Jaro Similarity. Nonetheless, there are still some small dissimilarities depending on the languages, it seems to works well for Spanish and  less well for German and Polish. We can also note that for Portuguese, if the preferred label is well translated ($\measureOne{}{}{}$, $\measureTwo{}{}{}$), it is less the case for the synonyms ($\measureThree{}{}{}$).

Then, the first-order links from Wikidata have also some satisfactory results, there are also dissimilarities between the languages. Especially, first-order links seem to work better than the average in French. Compared to second-order links, first-order links are always better and the decrease in quality between both is substantial. Some noise is probably added by the intermediate ontologies.


\subsection{Synonyms}
\label{subsection:synonyms}

\newcite{hailu_ontology_2014} suggests that synonyms play an important role in translation. Therefore, in addition to high-quality translation, we are also interested in a high number of synonyms. In our case, the synonyms are the different labels available for each language for Orphanet and Wikidata, and the translations of the English labels for Google Cloud Translation. We want to evaluate the richness of each methods in terms of numbers of synonyms. For a fair comparison, for each language we only work on the subset where the entities in Wikidata have at least one label in the evaluated language.

\begin{table}[!h]
\center
\begin{tabular}{|c|c|c|c|c|}

      \hline
      Lang & Orphanet & Wiki 1st & Wiki 1+2nd & GCT \\
      \hline
      EN & 2.3 & 5.8 & 166.77 & 2.3 \\
      \hline
      FR & 2.36 & 1.49 & 10.59 & 2.39 \\
      \hline
      DE & 2.56 & 1.84 & 5.93 & 2.65 \\
      \hline
      ES & 2.26 & 2.61 & 9.50 & 2.39 \\
      \hline
      PL & 2.54 & 2.01 & 6.88 & 2.65 \\
      \hline
      IT & 2.36 & 1.85 & 3.50 & 2.5 \\
      \hline
      PT & 1.62 & 1.60 & 2.40 & 2.41 \\
      \hline
      NL & 2.6 & 1.74 & 3.74 & 2.48 \\
      \hline
      CS & 2.2 & 1.74 & 1.71 & 2.13 \\
      \hline
\end{tabular}
\caption{Average number of labels in the different sources in function of the language. For Orphanet we only use the subset of entities linked to entities in Wikidata with at least one label in the studied language. For Google Cloud Translation, it is the translation of the English labels of Orphanet.}
\label{table:synonyms}
\center
\end{table}

Table \ref{table:synonyms} shows that generally Orphanet seems to have more synonyms than Wikidata when using first-order links only. And the fact GCT has more synonyms means that Orphanet has more labels in English than in other languages on the studied subset for majority language, except Dutch and Czech. Thus, this is not the case in English. For this language Wikidata is more diverse.

When using first and second-order links, the number of synonyms is much higher, especially for English. This is related to the fact that second-order links add many new relations. This new relations always have labels in English but not always habe labels in other languages.

\section{Discussion}

Regarding coverage, in terms of entities only, the coverage of first-order links is already high for Orphanet and Disease Ontology, respectively 84.9\% and 97.2\% (for English as, in our case, all the entities have English labels). The issue comes from the labels: even if Wikidata is multilingual, in our study we see that the information is mainly in English and French, but for the other studied languages the results are substantially worse. All the entities with a link have labels in English, more than half have labels in French and then for German, only around 20\% of the 8,870 linked entities in Wikidata have at least one label in German. The languages we study are among the most used languages in Wikipedia. Thus, it is already an important amount of entities that could have their labels translated from English to another of these languages. As Wikidata is a collaborative project, this number should only increase over time. 
Second-order links help a lot for languages other than English.  

Regarding quality, Google Cloud Translation is the best method. Compared to the results obtained by \newcite{silva_ontology-based_2015} on the translation of a subpart of MeSH in Portuguese, the quality of the label translations seems to have greatly improved. Then translations obtained through first-order links are not so distant from Google Cloud Translation. However, the quality of the translations obtained through second-order links has a substantial difference with the translation coming from first-order links. Thus, we can expect Google Cloud Translation to have an advantage as Orphanet is primarily maintained in English and French and then translated by experts to other languages. Even if Google Cloud Translation is not free, translating the entirety of the English labels of Orphanet would only cost around 16\$ with the pricing as of February 6, 2020.

For the synonyms, as Orphanet seems to have more labels in English than in the other languages, translating all the labels from English to the different languages allows having more synonyms than Orphanet in other languages.
Moreover, Wikidata is poorer in terms of synonyms than Orphanet except for English. This is interesting as Google Cloud Translation seems to perform good translations, and having more synonyms in English also means that if we translate them with Google Cloud Translation we could have also more synonyms in other languages. 
It is also important to note that Google Cloud Translation only provides one translation by label.
Second-order links also bring many more synonyms for all the languages, but especially for those which have a larger Wikidata.

\section{Conclusions and Future Work}

One of the limitations of this work concerns information that was not used. Especially in Orphanet and Wikidata, when an entity is linked to another ontology, there is additional information about the nature of the link, for example, whether it is an exact match or a more general entity. We did not use at all this information and it could be used to improve the links we create. Wikidata also contains more information about the entities than just the labels, e.g., \newcite{jiang_semantic_2013} extracts multilingual textual definitions.

We also focus our study on one type of biomedical entities, diseases. The results of this work may not be generalized to all types of entities. \newcite{hailu_ontology_2014} have found equivalent results for the translation of the Gene Ontology between English and German, but \newcite{silva_ontology-based_2015} did not find the same results on their partial translation of MeSH.


Another limitation is our study about synonyms. Having the maximum number of synonyms is useful for entity recognition and normalization. Thus, here we only have quantitatively studied the synonyms, and have not explored their quality and diversity. First- and second-order link extraction from Wikidata seems to be a good method to have more synonyms. A further assessment with an expert that could validate the synonyms could be interesting.

Furthermore, as we are interested in entity recognition, a low coverage on the ontology is not correlated with a low coverage for entities in a corpus. In \newcite{bretschneider_corpus-based_2014}, by only translating a small sub-part of an ontology they could improve the coverage of the entities in their corpus by a high margin. It will be interesting to verify this on a dataset on disease recognition.

To summarize, as of now, Google Cloud Translate seems to be the best way to translate an ontology about diseases. If the ontology does not have many synonyms, Wikidata could be a way to expand language-wise the ontology. Wikidata also contains other information about its entities which could be interesting, but have not been used in this study such as symptoms and links to Wikipedia pages.

\section{Bibliographical References}\label{reference}

\bibliographystyle{lrec}
\bibliography{extendont}


\end{document}